\documentclass[english,nocopyrightspace]{sigplanconf}
\usepackage{lmodern}

\usepackage[T1]{fontenc}
\usepackage[latin9]{inputenc}
\usepackage{verbatim}
\usepackage{amsthm}
\usepackage{amsmath}
\usepackage{amssymb}
\usepackage{graphicx}
\usepackage[numbers]{natbib}

\makeatletter
\numberwithin{equation}{section}
\numberwithin{figure}{section}
\theoremstyle{plain}
\newtheorem{thm}{\protect\theoremname}
\theoremstyle{definition}
\newtheorem{defn}[thm]{\protect\definitionname}
\theoremstyle{definition}
\newtheorem{example}[thm]{\protect\examplename}
\newcommand{\code}[1]{\texttt{#1}}

\@ifundefined{showcaptionsetup}{}{%
 \PassOptionsToPackage{caption=false}{subfig}}
\usepackage{subfig}
\makeatother

\usepackage{babel}
\usepackage{listings}
\providecommand{\definitionname}{Definition}
\providecommand{\examplename}{Example}
\providecommand{\theoremname}{Theorem}

\begin{document}

\title{Java Subtyping as an Infinite Self-Similar Partial Graph Product}

\authorinfo{Moez A. AbdelGawad}
{Informatics Research Institute, SRTA-City, Alexandria, Egypt}
{\texttt{moez@cs.rice.edu}}
\maketitle
\begin{abstract}
\global\long\def\pcgp{\ltimes}
Due to supporting variance annotations, such as wildcard types, the
structure of the subtyping relation in Java and other generic nominally-typed
OO programming languages is both interesting and intricate. In these
languages, the subtyping relation between \emph{ground} object types,
\emph{i.e.}, ones with no type variables, is the basis for defining
the full OO subtyping relation, \emph{i.e.}, that includes type variables.

As an ordering relation over the set of types, the subtyping relation
in object-oriented programming languages can always be represented
as a directed graph. In order to better understand some of the subtleties
of the subtyping relation in Java, in this paper we present how the
subtyping relation between ground Java types can be precisely constructed
using two new operations (a binary operation and a unary one) on directed
graphs. The binary operation we use, called a partial Cartesian graph
product, is similar in its essence to standard graph products and
group products. Its definition is based in particular on that of the
standard Cartesian graph product.

We believe the use of graph operations in constructing the ground
generic Java subtyping relation reveals some of the not-immediately-obvious
structure of the subtyping relation not only in Java but, more generally,
also in mainstream generic nominally-typed OO programming languages
such as C\#, Scala and Kotlin. Accordingly, we believe that describing
precisely how graph operations can be used to explicitly construct
the subtyping relation in these languages, as we do in this paper,
may significantly improve our understanding of features of the type
systems of these languages such as wildcard types and variance annotations,
and of the dependency of these features on nominal subtyping in nominally-typed
OOP.
\end{abstract}


\section{Introduction}

The addition of generics and wildcard types to Java~\citep{JLS05,JLS14}
made the subtying relation in Java elaborately intricate. Wildcard
types in Java express so-called \emph{usage-site} variance annotations~\citep{Torgersen2004}.
As~Torgersen et al. explain, supporting wildcard types
in Java causes the subtyping relation between generic types in Java
to be governed by three rules, namely:
\begin{itemize}
\item Covariant subtyping, which, for example, for a generic class\footnote{In this work we treat Java interfaces as abstract classes.}
\code{List} causes type \code{List<?~extends Integer>} to be a \emph{subtype}
of type \code{List<?~extends Number>} because type \code{Integer}
is a subtype of type \code{Number},
\item Contravariant subtyping, which causes type \code{List<?~super Integer>}
to be a \emph{supertype} of type \code{List<?~super Number>} because
type \code{Integer} is a subtype of type \code{Number}, and
\item Invariant subtyping, which causes type \code{List<Integer>} to be
\emph{unrelated} by subtyping to type \code{List<Number>} even\emph{
}when type \code{Integer} is a subtype of type \code{Number}.%

\end{itemize}
The subtyping relation in other industrial-strength generic nominally-typed
OOP languages such as C\#~\citep{CSharp2015}, Scala~\citep{Odersky14}
and Kotlin~\citep{Kotlin18} exhibits similar intricacy. C\#, Scala
and Kotlin support another form of variance annotations (called \emph{declaration-site}
variance annotations) as part of their support of generic OOP. Usage-site
and declaration-site variance annotations have a similar effect on
the structure of the generic subtyping relation in nominally-typed
OOP languages.

The introduction of variance annotations in mainstream OOP, even though
motivated by earlier research, has generated much additional interest
in researching generics and in having a good understanding of variance
annotations in particular. In this paper we augment this research
and improve on earlier research %
by presenting how a precise product-like graph operation can be used
to construct the subtyping relation in Java, exhibiting and making
evident in our construction the self-similarity in the definition
and construction of the relation, with the expectation that our construction
method will apply equally well to subtyping in other OO languages
such as C\#, Scala and Kotlin.

As such, this paper is structured as follows. In Section~\ref{sec:Background}
we discuss the intricacy and self-similarity of the subtyping relation
in Java, followed by an introduction to partial Cartesian graph products.
Then, in Section~\ref{sec:Constructing}, we define another new unary
graph operation (called the wildcards graph constructor), then we
present the formal construction of the subtyping relation in Java
using partial Cartesian graph products and the new unary operation.
In Section~\ref{sec:Examples} we present examples of the application
of our construction method that demonstrate how it works (in Appendix~\ref{sec:SageMath-Code}
we present SageMath code implementations for constructing our examples).
In Section~\ref{sec:Related-Work} we discuss some research that
is related to ours. We conclude in Section~\ref{sec:Discussion}
by discussing some conclusions we made and discussing some future
research that can build on our work.

\section{\label{sec:Background}Background}

In this section we give an example of how the subtyping relation in
a simple Java program can be constructed iteratively, based on the
type and subtype declarations in the program. We follow that by a
brief introduction to the partial Cartesian graph product operation
and a presentation of its formal definition.

\subsection{Iterative Construction of The Java Subtyping Relation}

To explore the intricacy of the subtyping relation in Java, we borrow
a simple example from~\citep{AbdelGawad2017a}. Let\textquoteright s
consider a simple generic class declaration. Assuming we have no classes
or types declared other than class \code{Object} (whose name we later
abbreviate to \code{O}), with a corresponding type that has the same
name, then the generic class declaration \code{class C<T> extends Object~\{\}}
results in a subclassing relation in which class \code{C} is a \emph{subclass}
of \code{O}.

For subtyping purposes in Java, it is useful to also assume the existence
of a special class \code{Null} (whose name we later abbreviate to
\code{N}) that is a subclass of all classes in the program, and whose
corresponding type, hence, is a subtype of all Java reference types.

Following the description of~\citep{AbdelGawad2017a}, the generic
subtyping relation in our Java program can be constructed iteratively,
based on the mentioned assumptions and the declaration of generic
class \code{C}. (As done by~\citep{AbdelGawad2017a}, we also assume
that a generic class takes only one type parameter, and that type
variables of all generic classes have type \code{O} as their upper
bound.)

Given that we have at least one generic class, namely \code{C}, we
should first note that, since generic types can be arbitrarily nested,
the generic subtyping relation will have an infinite number of types.
As such, to construct the infinite subtyping relation we go in iterations,
where we start with a finite first approximation to the relation then,
after each iteration, we get a step closer to the full infinite relation.
Since~\citep{AbdelGawad2017a} describes this informal construction
process in detail, we do not repeat less relevant details here. We
instead refer the interested reader to~AbdelGawad's
paper~\citep{AbdelGawad2017a,AbdelGawad2017}.

For the purposes of this paper however, the reader should appreciate
the intricacy of the generic subtyping relation in Java by noticing,
at least in a rough informal sense so far, the self-similarity that
is evident in the relation, where the subtyping relation between different
covariant types (those of the form `\code{C<?~extends~Type>}')
inside a result relation is the same as the relation between types
in the input relation. Contravariant subtyping results in an opposite
ordering relation (between different types of the form `\code{C<?~super~Type>}')
and invariant subtyping results in no relation (between different
types of the form `\code{C<Type>}').

To construct the final, most accurate version of the subtyping relation,
the process described above is continued ad infinitum. The purpose
of using partial Cartesian graph products to define the construction
method of the Java subtyping relation, as we do later in this paper,
is to formally present how each iteration in this construction process
can be precisely modeled mathematically, based on the informal intuitions
presented in the example above.%

\subsection{Partial Cartesian Graph Products ($\protect\pcgp$)}

Graph products are commonplace in computer science~\citep{Hammack2011}.
Graph products are often viewed as a convenient language with which
to describe structures. \citep{AbdelGawad2018a} presents a notion
of a \emph{partial }Cartesian graph product, which we use in constructing
the subtyping relation in generic nominally-typed OOP languages. A
full discussion of the partial Cartesian graph product operation,
denoted by $\pcgp$, is presented by~\citep{AbdelGawad2018a}. Here
we present a summary of its definition and of some of its properties
most relevant to constructing the generic OO subtyping relation.

\global\long\def\cgp{\square}
\global\long\def\setcp{\times}
\global\long\def\setdu{+}
\global\long\def\cgu{\dotplus}
\global\long\def\adj{\sim}

\begin{figure*}
\noindent \begin{centering}
\subfloat[$G_{1}$]{\noindent \protect\centering{}\protect\includegraphics[scale=0.4]{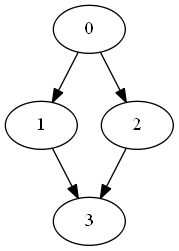}\protect}~~~~~~~~~~\subfloat[$G_{2}$]{\noindent \protect\centering{}\protect\includegraphics[scale=0.4]{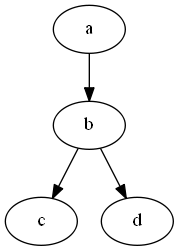}\protect}~~~~~~~~~~\subfloat[$G_{1}\protect\pcgp_{\{2,3\}}G_{2}$]{\noindent \protect\centering{}\protect\includegraphics[scale=0.4]{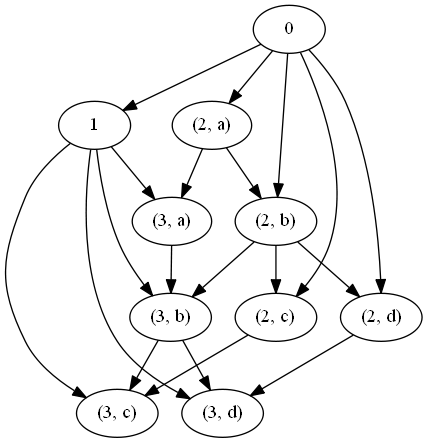}\protect}
\par\end{centering}

\protect\caption{\label{fig:Graphs-for-Illustrating}Illustrating the partial Cartesian
graph product operation $\protect\pcgp$}
\end{figure*}

\begin{defn}
\label{def:pcgp}(Partial Cartesian Graph Product, $\pcgp$). For
two directed graphs $G_{1}=\left(V_{1},E_{1}\right)$ and $G_{2}=\left(V_{2},E_{2}\right)$
where \end{defn}
\begin{itemize}
\item $V_{1}=V_{p}\setdu V_{n}$ such that $V_{p}$ and $V_{n}$ partition
$V_{1}$ (\emph{i.e.}, $V_{p}\subseteq V_{1}$ and $V_{n}=V_{1}\backslash V_{p}$),
\item $E_{1}=E_{pp}\setdu E_{pn}\setdu E_{np}\setdu E_{nn}$ such that $E_{pp}$,
$E_{pn}$, $E_{np}$, and $E_{nn}$ partition $E_{1}$,
\item $G_{p}=\left(V_{p},E_{pp}\right)$ and $G_{n}=\left(V_{n},E_{nn}\right)$
are two disjoint subgraphs of $G_{1}$ (the ones induced by $V_{p}$
and $V_{n}$, respectively, which guarantees that edges of $E_{pp}$
connect only vertices of $V_{p}$ and edges of $E_{nn}$ connect only
vertices of $V_{n}$), and $E_{pn}$ and $E_{np}$ connect vertices
from $V_{p}$ to $V_{n}$ and vice versa, respectively, and
\item $G_{2}$ is any directed graph (\emph{i.e.}, $G_{2}$, unlike $G_{1}$,
need not have some partitioning of its vertices and edges),
\end{itemize}
the\emph{ partial Cartesian graph product} of $G_{1}$ and $G_{2}$
relative to the set of vertices $V_{p}\subseteq V_{1}$ is the graph
\begin{equation}
G=G_{1}\pcgp_{V_{p}}G_{2}=\left(V,E\right)=G_{p}\cgp G_{2}\cgu G_{n}\label{eq:pcgp}
\end{equation}
 where
\begin{itemize}
\item $V=V_{p}\setcp V_{2}\setdu V_{n}$ ($\setcp$ and $\setdu$ are the
standard Cartesian set product and disjoint union operations),
\item $G_{p}\cgp G_{2}=\left(V_{p2},E_{p2}\right)$ is the standard Cartesian
graph product of $G_{p}$ and $G_{2}$, and,
\item for defining $E$, the operator $\cgu$ is defined (implicitly relative
to $G_{1}$) such that, if $\adj$ denotes adjacency, we have 
\[
\begin{cases}
\left(u{}_{1},v_{1}\right)\adj\left(u_{2},v_{2}\right)\in E & \textrm{if }\left(u{}_{1},v_{1}\right)\adj\left(u_{2},v_{2}\right)\in E_{p2}\\
\left(u_{1},v\right)\adj u_{2}\in E & \textrm{if }u_{1}\adj u_{2}\in E_{pn},v\in V_{2}\\
u_{1}\adj\left(u_{2},v\right)\in E & \textrm{if }u_{1}\adj u_{2}\in E_{np},v\in V_{2}\\
u_{1}\adj u_{2}\in E & \textrm{if }u_{1}\adj u_{2}\in E_{nn}
\end{cases}.
\]

\end{itemize}
Viewed abstractly, graph constructor $\pcgp$ is a binary graph operation
that takes as input two graphs $G_{1}$ and $G_{2}$ and is parameterized
also by a subset $V_{p}$ of the vertices $V_{1}$ of its first input
graph $G_{1}$. Informally, $\pcgp$ constructs a (standard) Cartesian
product of the subgraph induced by the given subset $V_{p}$ of vertices
with the second input graph $G_{2}$ and adds to this product the
non-product vertices of $G_{1}$ (\emph{i.e.}, $V_{1}\backslash V_{p}$)
appropriately connecting them to the product based on edges in $G_{1}$.
The example in Figure~\ref{fig:Graphs-for-Illustrating} helps demonstrate
the definition of $\pcgp$.

More details on the definition of $\pcgp$ and some of its properties
can be found in~\citep{AbdelGawad2018a}.

For using $\pcgp$ in constructing (the graph of) the generic OO subtyping
relation, the most interesting property of $\pcgp$ is that some vertices
in the first input graph $G_{1}$ do \emph{not} fully participate
in the product operation. This makes $\pcgp$ suited for constructing
the generic OO subtyping relation because some classes in a Java program
(where classes and types are mapped to vertices in the graphs of the
subclassing and subtyping relations) may not be generic classes (\emph{e.g.},
classes \code{Object} and \code{Null} are always non-generic), so
these classes do \emph{not }participate in the subtyping relation
with generic types (since they do not take type arguments to begin
with). We explain the construction method formally in the following
section.

\section{\label{sec:Constructing}Constructing The Java Subtyping Relation
Using $\protect\pcgp$}

\subsection{The Wildcards Graph Constructor ($\cdot^{\triangle}$)}

To construct the Java subtyping relation using $\pcgp$, we first
define a graph operator $\cdot^{\triangle}$ that constructs the graph
$S^{\triangle}$ whose vertices are all wildcard type arguments that
can be defined over the graph of a subtyping relation $S$, and whose
edges express the containment relation between these arguments. Informally,
the containment relation between wildcard type arguments is a very
simple relation, where we only have type argument `\code{T}' contained
in wildcard type argument `\code{?~<:~T}' (shorthand for `\code{?~extends~T}')
and contained in wildcard type argument `\code{?~:>~T}' (shorthand
for `\code{?~super~T}'), and also, by containment, we identify
the wildcard type arguments `\code{?}', `\code{?~<:~O}', and
`\code{?~:>~N}'. (If wildcard types are generalized to interval
types, as we do in~\citep{AbdelGawad2018c}, we have a fuller, more
elaborate containment relation).
\begin{defn}
(Triangle/Wildcards Graph, $\cdot^{\triangle}$) Formally, for a bipointed
graph $G=\left(V,E,v_{\top},v_{\bot}\right)$ (\emph{i.e.}, $G$ is
a graph with two distinguished ``source'' and ``sink'' vertices
$v_{\top},v_{\bot}\in V$, sometimes called top and bottom, or, for
our purposes, called \code{O} and \code{N}), the \emph{triangle}
graph (or, wildcards graph) $G^{\triangle}=\left(V_{\triangle},E_{\triangle}\right)$
of $G$ is defined as the reflexive transitive closure ($RTC$) of
the immediate (\emph{i.e.}, one-step) containment graph $G_{1}$ (\emph{i.e.},
$G^{\triangle}=RTC\left(G_{1}\right)$) where $G_{1}=\left(V_{\triangle},E_{1}\right)$
is defined as follows.\end{defn}
\begin{itemize}
\item $V_{\triangle}=V_{cov}\stackrel{\leftrightarrow}{\cup}V_{con}\stackrel{\leftrightarrow}{\cup}V_{inv}$,
such that $V_{cov}$, $V_{con}$, and $V_{inv}$ are three appropriately-labeled
``copies'' of $V$ corresponding to the three variant subtyping
rules (\emph{i.e.}, vertices in $V_{cov}$ are labeled with \code{?~<:~T}
for each label/type name \code{T} in $V$, while vertices in $V_{con}$
are labeled with \code{?~:>~T}, and vertices in $V_{inv}$ are labeled
with \code{T}---meaning that labels in $V_{inv}$ are exact copies
of the labels/type names in $V$), and such that the union-like operator
$\stackrel{\leftrightarrow}{\cup}$ identifies (\emph{i.e.}, coalesces)
the pair of vertices with labels/type names \code{?~<:~O} (denoting
all subtypes of \code{O}) and \code{?~:>~N} (denoting all supertypes
of \code{N}), the pair with \code{?~:>~O} (denoting all supertypes
of \code{O}) and \code{O}, and the pair with \code{?~<:~N} (denoting
all subtypes of \code{N}) and \code{N}. (Thus we have $|V_{\triangle}|=3*\left(|V|-1\right)$),
and,
\item for $E_{1}$, we have
\[
\begin{cases}
\mathtt{?\,<:\,T_{1}}\adj\mathtt{?\,<:\,T_{2}}\in E_{1} & \textrm{if }\mathtt{T_{1}}\adj\mathtt{T_{2}}\in E\\
\mathtt{?\,:>\,T_{2}}\adj\mathtt{?\,:>\,T_{1}}\in E_{1} & \textrm{if }\mathtt{T_{1}}\adj\mathtt{T_{2}}\in E\\
\mathtt{T}\adj\mathtt{?\,<:\,T}\in E_{1} & \mathtt{T}\in V\\
\mathtt{T}\adj\mathtt{?\,:>\,T}\in E_{1} & \mathtt{T}\in V
\end{cases}.
\]

\end{itemize}
As such, the graph operator $G^{\triangle}$ basically constructs
three ``copies''%
{} of the input graph $G$ (corresponding to the three variance subtyping
rules) and connects the vertices of these graphs based on the containment
relation\footnote{Hence the triangle symbol $\triangle$, where one copy---a ``side''
of the ``triangle''---is for the covariantly-ordered `\code{?~<:~T}'
wildcard type arguments, another side is for the oppositely/contravariantly-ordered
`\code{?~:>~T}' arguments, and the third, bottom side is for the
invariantly/flatly-ordered `\code{T}' arguments. See the graph
examples in Section~\ref{sec:Examples} for illustration, where green
edges correspond to the ``covariant side of the triangle'' and red
edges correspond to the ``contravariant side'', while the row/line
of vertices/types at the bottom (ones right above type \code{N},
with no interconnecting edges between them) correspond to the third
``invariant side''.}.

\subsection{Construction of The Java Subtyping Relation}

Based on the informal description of the construction method for the
Java subtyping relation and of the partial Cartesian graph product
constructor $\pcgp$ provided in Section~\ref{sec:Background}%
, we now define the graph $S$ of the subtyping relation of a particular
Java program as the solution of the recursive graph equation
\begin{equation}
S=C\pcgp_{C_{g}}S^{\triangle}\label{eq:S}
\end{equation}
where $C$ is the finite graph of the subclassing/inheritance relation
between classes of the program, and $C_{g}$ is the set of generic
classes of the program (a subset of classes in $C$).

Equation~(\ref{eq:S}) can be solved for $S$ (as the \emph{least}
fixed point of Equation~(\ref{eq:S})) iteratively, using the equation
\begin{equation}
S_{i+1}=C\pcgp_{C_{g}}S_{i}^{\triangle}\label{eq:Si}
\end{equation}
where the $S_{i}$ are finite successive better approximations of
the infinite relation $S$, and $S_{0}^{\triangle}$ is an appropriate
initial graph of the containment relation (which we take as the graph
with one vertex, having the default wildcard type argument `\code{?}',
standing for \code{?~<:~O}, as its only vertex, and having no containment
relation edges%
).\footnote{We conjecture that the \emph{greatest} fixed point of Equation~(\ref{eq:S})
may be useful in modeling ``F-bounded generics'', given the strong
connection between gfps (greatest fixed points) and coalgebras. The
study of coalgebras seems to be the area of universal algebra (and
category theory) on which the theory of F-bounded polymorphism (and
F-bounded generics) is based. We do not explore this point any further
in this paper however.}

\subsection{Self-Similarity and The Role of Nominal-Typing}

Equation~(\ref{eq:Si}) formally and succinctly describes the construction
method of the generic Java subtyping relation between ground Java
reference types. The self-similarity in the Java subtyping relation
can now be clearly seen to result from the fact that the second factor
in the partial product defining $S$ (\emph{i.e.},\emph{ }the wildcards
graph $S^{\triangle},$ the graph of the containment relation between
wildcard type arguments) is derived iteratively, in all but the first
iteration, from the first factor of the product (\emph{i.e.}, from
$C$, the subclassing relation).

Noting that subclassing/type inheritance is an inherently nominal
notion in OOP (\emph{i.e.}, is always defined using class names, to
express that corresponding named classes preserve inherited behavioral
contracts associated with the names), the observation of the dependency
of $S$ on $C$ makes it evident that the dependency of the OO subtyping
relation on the nominal subclassing/inheritance relation in mainstream
nominally-typed OO programming languages (such as Java, C\#, C++,
Scala and Kotlin, as discussed by~\citep{NOOPsumm,AbdelGawad14,InhSubtyNWPT13,AbdelGawad2015})
has strongly continued after generic types were added to these languages,
further illustrating the value of nominal typing in mainstream OOP
languages and the influence and effects of nominal typing on their
type systems.

Further, it should be noted that the properties of the partial Cartesian
graph product (as presented by~\citep{AbdelGawad2018a} and summarized
above) imply that non-generic OOP is a special case of generic OOP,
which is a fact that is intuitively clear to OO software developers.
In particular, the property of $\pcgp$ that for $C_{g}=\phi$ (\emph{i.e.},
if no classes in $C$ are generic) the result of the partial product
is equal to the left factor of the partial product makes Equation~(\ref{eq:S})
become 
\[
S=C,
\]
which just expresses the identification of the subtyping relation
($S$) with the subclassing/type inheritance relation ($C$), as is
well-known to be the case in non-generic nominally-typed OOP (again,
see~\citep{NOOPsumm,AbdelGawad14,InhSubtyNWPT13,AbdelGawad2015}).

\medskip{}

To strengthen our understanding of the formal construction of the
Java subtyping relation $S$, we can go on to discuss some of the
properties of the graphs $S_{i}$ approximating $S$, particularly
their size, order and element rank properties. Given that there are
formulae for (bounds on) the size and order of the product graph constructed
by $\pcgp$~\citep{AbdelGawad2018a} and of the triangle graph constructed
by $\triangle$ in terms of the sizes and orders of their input graphs,
(bounds on) the size and order of a graph $S_{i+1}$ can be computed
in terms of the size and order of $C$, $C_{g}$ and $S_{i}$, recursively
going down to $S_{0}^{\triangle}$ (with size 1 and order 0, \emph{i.e.},
one vertex and zero edges). For the sake of brevity, though, we do
not present equations%
{} of the size and order of $S_{i}$ in this paper\footnote{Even though an interesting mental exercise (and programming exercise,
given that the values given by the equations can be verified for some
sample graphs using SageMath code that builds on the code we present
in Appendix~\ref{sec:SageMath-Code}), we currently see the equations
as being interesting more as a mathematical curiosity than of them
having much practical or theoretical value.}\textsuperscript{,}\footnote{\label{fn:order}As the equations presented in~\citep{AbdelGawad2018a}
demonstrate, the reader may also like to note that coalescence of
multiedges---for graphs that model posets such as the OO subtyping
relation (where multiedges are meaningless)---and transitive reduction---used
to present graphs of subtyping as Hasse diagrams---make the equations
for computing the orders of constructed graphs in particular not immediately
straightforward or simple. In fact there is no known formula for computing
the order of the transitive reduction of a general graph in terms
of only the order and size of the graph. The same inexistence of formulae
applies to the order of the transitive closure of a graph, the number
of paths of a graph, and, seemingly, also the number of `graph intervals'~\citep{AbdelGawad2018c}
of a graph.}, and proceed to illustrating examples instead.

\section{\label{sec:Examples}Examples of Constructing The Java Subtyping
Relation}

In this section we present examples of how the generic Java subtyping
relation between ground reference types with wildcard type arguments
can be iteratively constructed. %
To decrease clutter, given that OO subtyping is a \emph{transitive}
relation, the transitive reduction of the subtyping graphs is presented
in the examples below. Also, we use colored edges of graphs to indicate
the self-similarity of the Java subtyping relation, where green edges
correspond to subtyping due to the covariant subtyping rule, while
red edges correspond to subtyping due to the contravariant subtyping
rule. (Note also, as we explained in the description of graph operator
$\cdot^{\triangle}$, that types \code{C<?>} and \code{C<?~<:~O>}
and \code{C<?~:>~N>} are different expressions of the same type,
\emph{i.e.}, are identified. The same applies to types \code{C<O>}
and \code{C<?~:>~O>}, as well as types \code{C<N>} and \code{C<?~<:~N>}).
\begin{example}
Consider the Java class declaration

\code{\textbf{class}~C<T>~\{\}}.

\noindent The graphs in Figure~\ref{fig:Illustrating-the-use} illustrate
the first three iterations of the construction of the subtyping relation
$S_{1}$ corresponding to this declaration.\footnote{\noindent The subscript $_{1}$ is used in $S_{1}$ and following
examples to denote the example index. We use double-indexing to additionally
refer to iteration indices, as in $S_{11}$, for example, which denotes
the first version of $S_{1}$ as defined by the first iteration of
the iterative construction method.} In Figure~\ref{fig:C2}, so as to shorten names of types in $S_{13}$,
we use \code{T1} to \code{T6} to stand for types of $S_{12}$ other
than \code{O} and \code{N}.

\noindent 
\begin{figure*}
\noindent \begin{centering}
\subfloat[$C_{1}$]{\noindent \protect\centering{}\protect\includegraphics[scale=0.5]{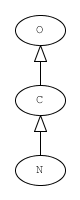}\protect}~~~~~\subfloat[$S_{11}=C_{1}\protect\pcgp_{\{\mathtt{C}\}}S_{10}^{\triangle}$]{\noindent \protect\centering{}~~~~~~~\protect\includegraphics[scale=0.5]{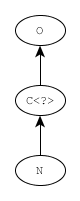}~~~~~~~\protect}~~~~~\subfloat[$S_{12}=C_{1}\protect\pcgp_{\{\mathtt{C}\}}S_{11}^{\triangle}$]{\noindent \protect\centering{}\protect\includegraphics[scale=0.4]{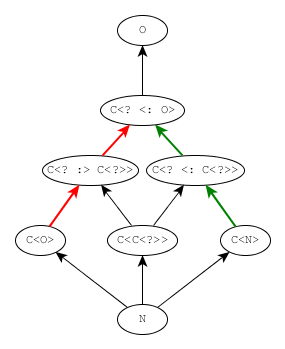}\protect}
\par\end{centering}

\noindent \begin{centering}
\subfloat[\label{fig:C2}$S_{13}=C_{1}\protect\pcgp_{\{\mathtt{C}\}}S_{12}^{\triangle}$]{\noindent \protect\centering{}\protect\includegraphics[scale=0.45]{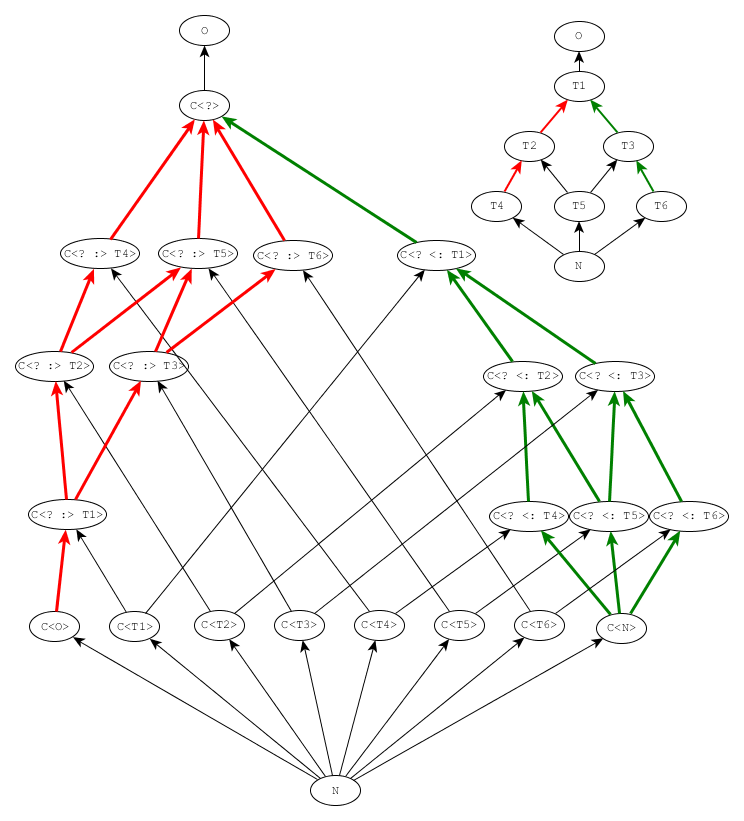}\protect}
\par\end{centering}

\protect\caption{\label{fig:Illustrating-the-use}Constructing generic OO subtyping
using $\protect\pcgp$}
\end{figure*}

\end{example}
\smallskip{}

\begin{figure*}
\begin{centering}
\subfloat[$C_{2}$]{\noindent \protect\centering{}\protect\includegraphics[scale=0.5]{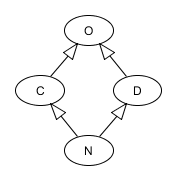}\protect}~~~~~~~~~~\subfloat[$S_{21}=C_{2}\protect\pcgp_{\{\mathtt{D}\}}S_{20}^{\triangle}$]{\protect\centering{}\protect\includegraphics[scale=0.5]{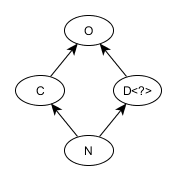}\protect}
\par\end{centering}

\centering{}\subfloat[$S_{22}=C_{2}\protect\pcgp_{\{\mathtt{D}\}}S_{21}^{\triangle}$]{\protect\centering{}\protect\includegraphics[scale=0.5]{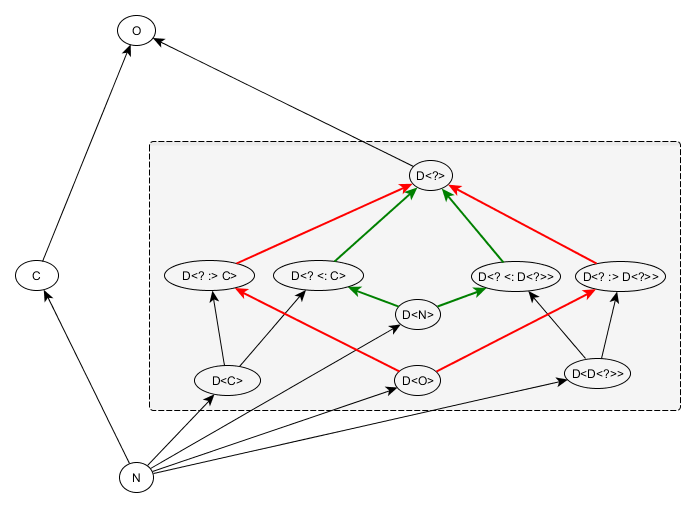}\protect}\protect\caption{\label{fig:Illustrating-the-use-1}Constructing generic OO subtyping
using $\protect\pcgp$}
\end{figure*}

\begin{example}
Consider the two Java class declarations

\code{\textbf{class}~C~\{\} } \emph{\code{\emph{//}} }class \code{C}
is non-generic

\code{\textbf{class}~D<T>~\{\}}.

\noindent The graphs in Figure~\ref{fig:Illustrating-the-use-1}
illustrate the first two iterations of the construction of the subtyping
relation $S_{2}$ corresponding to these declarations. (We do not
present graphs of $S_{x3}$ in this and later examples below, due
to the large size of these graphs).
\end{example}

\smallskip{}

\begin{figure*}
\noindent \begin{centering}
\subfloat[$C_{3}$]{\noindent \protect\centering{}\protect\includegraphics[scale=0.5]{SubclassingCD}\protect}~~~~~~~~~\subfloat[$S_{31}=C_{3}\protect\pcgp_{\{\mathtt{C},\mathtt{D}\}}S_{30}^{\triangle}$]{\noindent \protect\centering{}~~~~\protect\includegraphics[scale=0.5]{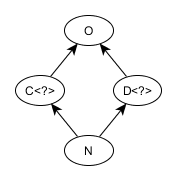}~~~~\protect}
\par\end{centering}

\centering{}\subfloat[$S_{32}=C_{3}\protect\pcgp_{\{\mathtt{C},\mathtt{D}\}}S_{31}^{\triangle}$]{\noindent \protect\centering{}\protect\includegraphics[scale=0.35]{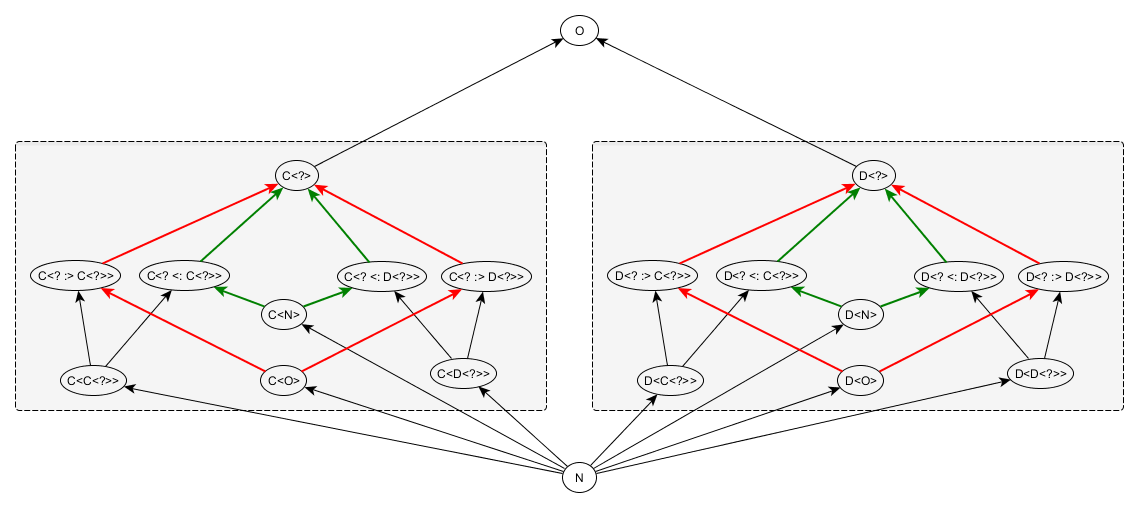}\protect}\protect\caption{\label{fig:Illustrating-the-use-2}Constructing generic OO subtyping
using $\protect\pcgp$}
\end{figure*}

\begin{example}
Consider the two Java class declarations

\code{\textbf{class}~C<T>~\{\}}

\code{\textbf{class}~D<T>~\{\}}.

\noindent The graphs in Figure~\ref{fig:Illustrating-the-use-2}
illustrate the first two iterations of the construction of the subtyping
relation $S_{3}$ corresponding to these declarations.
\end{example}

\smallskip{}

\begin{figure*}
\noindent \begin{centering}
\subfloat[$C_{4}$]{\noindent \protect\centering{}\protect\includegraphics[scale=0.5]{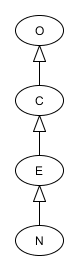}\protect}~~~~~~~~~\subfloat[$S_{41}=C_{4}\protect\pcgp_{\{\mathtt{C},\mathtt{E}\}}S_{40}^{\triangle}$]{\noindent \protect\centering{}~~~~~~~~~\protect\includegraphics[scale=0.5]{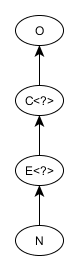}~~~~~~~~~\protect}
\par\end{centering}

\noindent \centering{}\subfloat[$S_{42}=C_{4}\protect\pcgp_{\{\mathtt{C},\mathtt{E}\}}S_{41}^{\triangle}$]{\noindent \protect\centering{}\protect\includegraphics[scale=0.44]{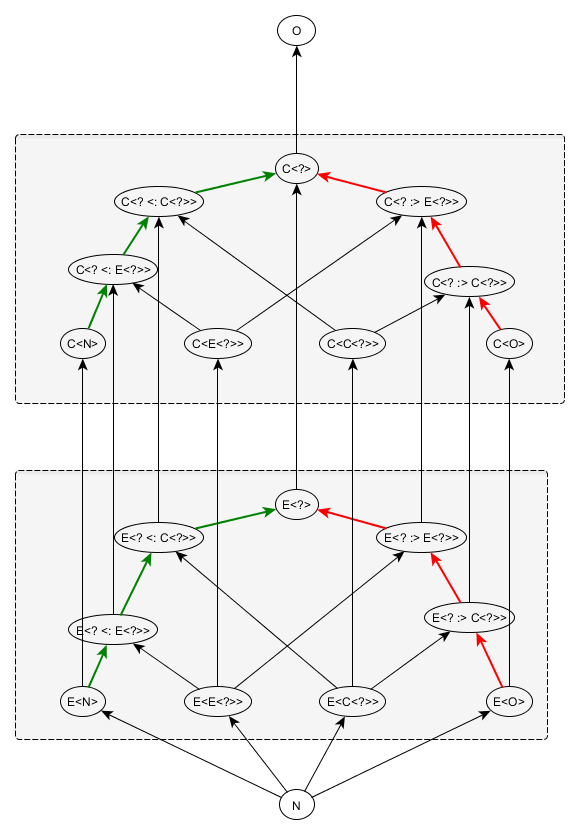}\protect}\protect\caption{\label{fig:Illustrating-the-use-3}Constructing generic OO subtyping
using $\protect\pcgp$}
\end{figure*}

\begin{example}
Consider the two Java class declarations

\code{\textbf{class}~C<T>~\{\}}

\code{\textbf{class}~E<T> \textbf{extends} C<T>~\{\}}. 

\noindent The graphs in Figure~\ref{fig:Illustrating-the-use-3}
illustrate the first two iterations of the construction of the subtyping
relation $S_{4}$ corresponding to these declarations. 
\end{example}

\smallskip{}

\begin{figure*}
\noindent \begin{centering}
\subfloat[$C_{5}$]{\noindent \protect\centering{}\protect\includegraphics[scale=0.6]{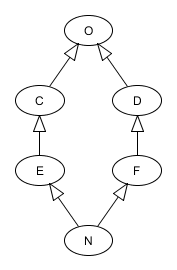}\protect}~~~~~~~~~\subfloat[$S_{51}=C_{5}\protect\pcgp_{\{\mathtt{F}\}}S_{50}^{\triangle}$]{\noindent \protect\centering{}\protect\includegraphics[scale=0.6]{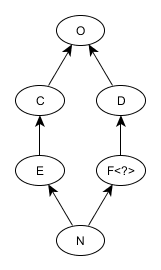}\protect}
\par\end{centering}

\noindent \centering{}\subfloat[$S_{52}=C_{5}\protect\pcgp_{\{\mathtt{F}\}}S_{51}^{\triangle}$]{\noindent \protect\centering{}\protect\includegraphics[scale=0.55]{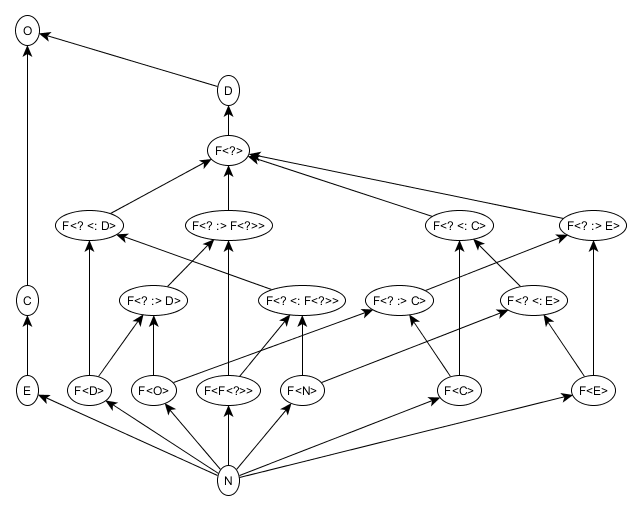}\protect}\protect\caption{\label{fig:Illustrating-the-use-4}Constructing generic OO subtyping
using $\protect\pcgp$ (automatic layout by yEd)}
\end{figure*}

\begin{example}
Consider the four Java class declarations

\code{\textbf{class}~C~\{\}}

\code{\textbf{class}~E \textbf{extends} C~\{\}}

\code{\textbf{class}~D~\{\}}

\code{\textbf{class}~F<T> \textbf{extends} D~\{\}}.

\noindent The graphs in Figure~\ref{fig:Illustrating-the-use-4}
illustrate the first two iterations of the construction of the subtyping
relation $S_{5}$ corresponding to these declarations. (Readers are
invited to find out the subgraphs of $S_{52}$ that are similar---\emph{i.e.},
isomorphic---to $S_{51}$, and, at least mentally in their minds,
to layout $S_{52}$ accordingly.)
\end{example}

(In Appendix~\ref{sec:SageMath-Code} we present SageMath~\citep{Stein2017}
code that helped us---and can help the readers---in producing some
of the diagrams presented in this paper.)

\section{\label{sec:Related-Work}Related Work}

Using a graph product to construct the generic OO subtyping relation
seems to be a new idea, with no similar prior work. We already mentioned,
however, the earlier work of~\citep{AbdelGawad2017a} that uses category
theoretic tools (namely operads) to model generic OO subtyping, which
is work that has paved the way for the work we present in this paper.

The addition of generics to Java has motivated much earlier research
on generic OOP and also on the type safety of Java and similar languages.
Much of this research was done before generics were added to Java.
For example, the work of~\citep{Bank96,Bracha98,Corky98} was mostly
focused on researching OO generics, while the work of~\citep{drossopoulou99,flatt99}
was focused on type safety.

Some research on generics was also done after generics were added
to Java (\emph{e.g.},~\citep{Zhang:2015:LFO:2737924.2738008,AbdelGawad2016c,Grigore2017,AbdelGawad2017b}).
However, Featherweight Java/Featherweight Generic Java (FJ/FGJ)~\citep{FJ/FGJ}
is probably the most prominent work done on the type safety of Java,
including generics. Variance annotations and wildcard types were not
put in consideration in the construction of the operational model
of generic nominally-typed OOP presented by~\citep{FJ/FGJ} however.

Separately, probably as the most complex feature of Java generics,
the addition of \textquotedblleft wildcards\textquotedblright{} (\emph{i.e.},
wildcard type arguments) to Java (in the work of~\citep{Torgersen2004},
which is based on the earlier research by~\citep{Igarashi02onvariance-based})
also generated some research that is particularly focused on modeling
wildcards and variance annotations~\citep{MadsTorgersen2005,KennedyDecNomVar07,Cameron2008,Summers2010,Tate2011,Tate2013,Greenman2014}.
This substantial work points to the need for more research on wildcard
types and generic OOP.

\section{\label{sec:Discussion}Discussion and Future Work}

\global\long\def\JSO{\mathcal{JSO}}
In this paper we demonstrate, much more precisely than was done by~\citep{AbdelGawad2017a},
how the graph of the Java subtyping relation between ground Java reference
types can be constructed as an infinite self-similar partial Cartesian
graph product. The simple construction method we presented in this
paper nicely captures some of the main features of the generic subtyping
relation in Java and similar OO languages, particularly the details
of the self-similarity of the relation. Based on our development of
a notion of a partial Cartesian graph product and the earlier development
of the outline of the $\JSO$ operad (by~\citep{AbdelGawad2017a})
for use in modeling the generic Java subtyping relation (both of which
particularly reveal the intricate self-similarity of the relation)
we strongly believe that using more mathematical tools from category
theory (such as operads) and from graph theory (such as partial graph
products) is very likely to be the key to having a better understanding
of complex features of programming languages such as wildcard types
and generics.

In agreement with the detailed explanation of~\citep{AbdelGawad2017a},
in our opinion the most important reason for obscuring the self-similarity
of the generic subtyping relation in Java, and the exact details of
the intricacy of its self-similarity, is thinking about the subtyping
relation in structural-typing terms rather than nominal-typing ones.
Although the polymorphic structural subtyping relation, with variance
annotations, may exhibit some form of self-similarity that is similar
to the one we demonstrate for Java, it should be noted that, as explained
by~\citep{NOOPsumm,InhSubtyNWPT13,AbdelGawad14,AbdelGawad2015},
nominal typing in OO languages such as Java, C\#, Kotlin and Scala
causes the full identification of (\emph{i.e.}, one-to-one correspondence
between) type/contract inheritance and nominal subtyping in non-generic
OOP. Such a simple and strong connection between type inheritance
and subtyping does \emph{not} exist when thinking about the OO subtyping
relation in structural typing terms. Based on the discussion of the
source of self-similarity in the Java subtyping relation that we present
in Section~\ref{sec:Constructing}, it seems to us that not making
this observation, keeping instead the subtyping relation separate
and independent from the inheritance relation, makes it harder to
see the self-similarity of generic nominal subtyping, its intricacies,
and its fundamental dependency on the subclassing/inheritance relation.

Having said that, more work is needed, however, to model Java and
generic nominal subtyping more accurately. In particular, in this
paper we do not model bounded type variables (other than those upper-bounded
with type \code{Object}). An immediate extension of our work, that
gets us closer to modeling bounded type variables (with both lower
and upper bounds), is to construct a more general Java subtyping relation
that uses interval types instead of wildcard types (we conjecture
that lower bounds on type variables, which are unsupported so far
in Java, will mesh well with interval types, which~\citep{AbdelGawad2018c}
introduces).

Other investigations that can build on the work we present here are
to construct the Java subtyping relation with less restrictions/assumptions,
such as allowing more complex inheritance relations between generic
types\footnote{So as to allow, for example, generic class declarations such as \code{class C<T> extends D<E<T>\textcompwordmark{}>~\{\}}
and \code{class C<T> extends D<E<F<T>\textcompwordmark{}>~\{\}} (assuming
the simple obvious declarations of generic classes \code{D}, \code{E},
and \code{F}). We believe these type inheritance declarations can
be handled in (a simple extension of) our Java subtyping construction
method by introducing instantiations of class \code{C} one and two
iterations \emph{after} (\emph{i.e.}, will be of a higher rank than)
generic instantiations of classes \code{E} and \code{F} respectively.
Instantiation of a generic class \code{C} in its own supertype (\emph{e.g.},
as in the Java class declaration \code{class C<T> extends D<C<T>\textcompwordmark{}>~\{\}})
presents a circularity problem however. (Although similar in flavor,
and related, this issue is separate from the notion of ``F-bounded
generics'').} and allowing multiple type arguments. To analyze the full Java subtyping
relation, \emph{i.e.}, between ground and non-ground reference types,
type variables may also be included in the construction of the Java
subtyping relation. We suggest in~\citep{AbdelGawad2016a,AbdelGawad2016c,AbdelGawad2017a,AbdelGawad2017b}
how this might be done, but we leave the actual work to future work.

\bibliographystyle{plain}

\begin{thebibliography}{10}
	
	\bibitem{CSharp2015}
	C\# language specification, version 5.0.
	\newblock http://msdn.microsoft.com/vcsharp, 2015.
	
	\bibitem{Kotlin18}
	Kotlin language documentation, v. 1.2.
	\newblock http://www.kotlinlang.org, 2018.
	
	\bibitem{NOOPsumm}
	Moez~A. AbdelGawad.
	\newblock A domain-theoretic model of nominally-typed object-oriented
	programming.
	\newblock {\em Electronic Notes in Theoretical Computer Science (ENTCS)},
	301:3--19, 2014.
	
	\bibitem{AbdelGawad2016c}
	Moez~A. AbdelGawad.
	\newblock Towards an accurate mathematical model of generic nominally-typed
	{OOP} (extended abstract).
	\newblock {\em {arXiv:1610.05114 [cs.PL]}}, 2016.
	
	\bibitem{AbdelGawad2016a}
	Moez~A. AbdelGawad.
	\newblock Towards understanding generics.
	\newblock Technical report, {arXiv:1605.01480 [cs.PL]}, 2016.
	
	\bibitem{AbdelGawad2015}
	Moez~A. AbdelGawad.
	\newblock Why nominal-typing matters in {OOP}.
	\newblock {\em Preprint available at http://arxiv.org/abs/1606.03809}, 2016.
	
	\bibitem{AbdelGawad2017b}
	Moez~A. AbdelGawad.
	\newblock Novel uses of category theory in modeling {OOP} (extended abstract).
	\newblock {\em Accepted at The Nordic Workshop on Programming Theory (NWPT'17),
		Turku, Finland, November 1-3, 2017. (Full version available at arXiv.org:
		1709.08056 [cs.PL])}, 2017.
	
	\bibitem{AbdelGawad2017a}
	Moez~A. AbdelGawad.
	\newblock Towards a {J}ava subtyping operad.
	\newblock {\em Proceedings of FTfJP'17, Barcelona, Spain, June 18-23, 2017},
	2017.
	
	\bibitem{AbdelGawad2017}
	Moez~A. AbdelGawad.
	\newblock Towards a {J}ava subtyping operad (extended version).
	\newblock {\em Preprint available at http://arxiv.org/abs/1706.00274}, 2017.
	
	\bibitem{AbdelGawad2018a}
	Moez~A. AbdelGawad.
	\newblock Partial cartesian graph product (and its use in modeling {J}ava
	subtyping).
	\newblock {\em Available as arXiv preprint at http://arxiv.org/abs/1805.07155},
	2018.
	
	\bibitem{AbdelGawad2018c}
	Moez~A. AbdelGawad.
	\newblock Towards taming {J}ava wildcards and extending {J}ava with interval
	types.
	\newblock {\em Available as arXiv preprint at http://arxiv.org/abs/1805.10931},
	2018.
	
	\bibitem{AbdelGawad14}
	Moez~A. AbdelGawad and Robert Cartwright.
	\newblock {NOOP}: A domain-theoretic model of nominally-typed object-oriented
	programming.
	\newblock {\em Available as arXiv preprint at http://arxiv.org/abs/1801.06793},
	2015.
	
	\bibitem{Bank96}
	Joseph~A. Bank, Barbara Liskov, and Andrew~C. Myers.
	\newblock Parameterized types and {J}ava.
	\newblock Technical report, 1996.
	
	\bibitem{Bracha98}
	Gilad Bracha, Martin Odersky, David Stoutamire, and Philip Wadler.
	\newblock Making the future safe for the past: Adding genericity to the {J}ava
	programming language.
	\newblock In Craig Chambers, editor, {\em ACM Symposium on Object-Oriented
		Programming: Systems, Languages and Applications (OOPSLA)}, volume~33, pages
	183--200, Vancouver, BC, October 1998. ACM, ACM SIGPLAN.
	
	\bibitem{Cameron2008}
	Nicholas Cameron, Sophia Drossopoulou, and Erik Ernst.
	\newblock A model for {J}ava with wildcards.
	\newblock In {\em ECOOP'08}, 2008.
	
	\bibitem{InhSubtyNWPT13}
	Robert Cartwright and Moez~A. AbdelGawad.
	\newblock Inheritance \emph{Is} subtyping (extended abstract).
	\newblock In {\em The 25\textsuperscript{th} Nordic Workshop on Programming
		Theory (NWPT)}, Tallinn, Estonia, 2013.
	
	\bibitem{Corky98}
	Robert Cartwright and Jr. Steele, Guy~L.
	\newblock Compatible genericity with run-time types for the {J}ava programming
	language.
	\newblock In Craig Chambers, editor, {\em ACM Symposium on Object-Oriented
		Programming: Systems, Languages and Applications (OOPSLA)}, volume~33, pages
	201--215, Vancouver, BC, October 1998. ACM, ACM SIGPLAN.
	
	\bibitem{drossopoulou99}
	Sophia Drossopoulou, Susan Eisenbach, and Sarfraz Khurshid.
	\newblock Is the {J}ava type system sound?
	\newblock {\em TAPOS}, 5(1):3--24, 1999.
	
	\bibitem{flatt99}
	Matthew Flatt, Shriram Krishnamurthi, and Matthias Felleisen.
	\newblock A programmer's reduction semantics for classes and mixins.
	\newblock In {\em Formal syntax and semantics of Java}, pages 241--269.
	Springer, 1999.
	
	\bibitem{JLS05}
	James Gosling, Bill Joy, Guy Steele, and Gilad Bracha.
	\newblock {\em The {J}ava Language Specification}.
	\newblock Addison-Wesley, 2005.
	
	\bibitem{JLS14}
	James Gosling, Bill Joy, Guy Steele, Gilad Bracha, and Alex Buckley.
	\newblock {\em The {J}ava Language Specification}.
	\newblock Addison-Wesley, 2014.
	
	\bibitem{Greenman2014}
	Ben Greenman, Fabian Muehlboeck, and Ross Tate.
	\newblock Getting f-bounded polymorphism into shape.
	\newblock In {\em {PLDI} '14: Proceedings of the 2014 {ACM} {SIGPLAN}
		conference on Programming Language Design and Implementation}, 2014.
	
	\bibitem{Grigore2017}
	Radu Grigore.
	\newblock {J}ava generics are {T}uring complete.
	\newblock In {\em Proceedings of the 44th ACM SIGPLAN Symposium on Principles
		of Programming Languages}, POPL 2017, pages 73--85, New York, NY, USA, 2017.
	ACM.
	
	\bibitem{Hammack2011}
	Richard Hammack, Wilfried Imrich, and Sandi Klavzar.
	\newblock {\em Handbook of Product Graphs}.
	\newblock CRC Press, second edition edition, 2011.
	
	\bibitem{FJ/FGJ}
	Atsushi Igarashi, Benjamin~C. Pierce, and Philip Wadler.
	\newblock {F}eatherweight {J}ava: A minimal core calculus for {J}ava and {GJ}.
	\newblock {\em ACM Transactions on Programming Languages and Systems},
	23(3):396--450, May 2001.
	
	\bibitem{Igarashi02onvariance-based}
	Atsushi Igarashi and Mirko Viroli.
	\newblock On variance-based subtyping for parametric types.
	\newblock In {\em In ECOOP}, pages 441--469. Springer-Verlag, 2002.
	
	\bibitem{KennedyDecNomVar07}
	Andrew~J. Kennedy and Benjamin~C. Pierce.
	\newblock On decidability of nominal subtyping with variance.
	\newblock In {\em International Workshop on Foundations and Developments of
		Object-Oriented Languages (FOOL/WOOD)}, 2007.
	
	\bibitem{Odersky14}
	Martin Odersky.
	\newblock The {S}cala language specification, v. 2.9.
	\newblock http://www.scala-lang.org, 2014.
	
	\bibitem{Stein2017}
	William Stein.
	\newblock Sagemath 8.1.
	\newblock {\em http://www.sagemath.org}, 2017.
	
	\bibitem{Summers2010}
	Alexander~J. Summers, Nicholas Cameron, Mariangiola Dezani-Ciancaglini, and
	Sophia Drossopoulou.
	\newblock Towards a semantic model for {J}ava wildcards.
	\newblock {\em 10\textsuperscript{th} Workshop on Formal Techniques for
		{J}ava-like Programs}, 2010.
	
	\bibitem{Tate2013}
	Ross Tate.
	\newblock Mixed-site variance.
	\newblock In {\em {FOOL} '13: Informal Proceedings of the 20th International
		Workshop on Foundations of Object-Oriented Languages}, 2013.
	
	\bibitem{Tate2011}
	Ross Tate, Alan Leung, and Sorin Lerner.
	\newblock Taming wildcards in {J}ava's type system.
	\newblock {\em PLDI'11, June 4--8, 2011, San Jose, California, USA.}, 2011.
	
	\bibitem{MadsTorgersen2005}
	Mads Torgersen, Erik Ernst, and Christian~Plesner Hansen.
	\newblock Wild {FJ}.
	\newblock In {\em Foundations of Object-Oriented Languages}, 2005.
	
	\bibitem{Torgersen2004}
	Mads Torgersen, Christian~Plesner Hansen, Erik Ernst, Peter von~der Ah\'e,
	Gilad Bracha, and Neal Gafter.
	\newblock Adding wildcards to the {J}ava programming language.
	\newblock In {\em SAC}, 2004.
	
	\bibitem{Zhang:2015:LFO:2737924.2738008}
	Yizhou Zhang, Matthew~C. Loring, Guido Salvaneschi, Barbara Liskov, and
	Andrew~C. Myers.
	\newblock Lightweight, flexible object-oriented generics.
	\newblock In {\em Proceedings of the 36th ACM SIGPLAN Conference on Programming
		Language Design and Implementation}, PLDI 2015, pages 436--445, New York, NY,
	USA, 2015. ACM.
	
\end{thebibliography}

\appendix

\section{\label{sec:SageMath-Code}SageMath Code}

In this appendix we present the SageMath~\citep{Stein2017} code
that we used to help produce some of the graph examples presented
in this paper. The code presented here is not optimized for speed
of execution but rather for clarity and simplicity of implementation.

\begin{lstlisting}[language=Python,basicstyle={\small\ttfamily},frame=lines]
TopCls = 'O'
BotCls = 'N'

WLP = '<' # LeftParen for wildcard type args
WRP = '>' # RightParen for wildcard type args

ExtStr = ' <: '
SupStr = ' :> '

W = '?' # Wildcard
WExt = W+ExtStr
WSup = W+SupStr

#Construct Type Arguments (TAs)
def TAs(S):
  CovTA=S.copy()
  CovTA.relabel(lambda T: W if (T==TopCls) else
                          (BotCls if (T==BotCls)
                           else WExt+T))

  ConTA=S.copy()
  ConTA.reverse_edges(ConTA.edges())
  # all of them, due to contravariant subtyping
  ConTA.relabel(lambda T: W if (T==BotCls) else
                          (TopCls if (T==TopCls)
                           else WSup+T))

  InvTA=S.copy()
  InvTA.delete_edges(InvTA.edges())
  # all of them, due to invariant subtyping

  TA=CovTA.union(ConTA).union(InvTA)
  # earlier relabelling helps identify type args.

  # Add subtyping edges from InvTA to corrspndng
  # type args in CovTA and ConTA.
  MinON = InvTA.copy()
  MinON.delete_vertex(TopCls)
  MinON.delete_vertex(BotCls)
  TA.add_edges([(T,WExt+T) for T in MinON])
  TA.add_edges([(T,WSup+T) for T in MinON])

  TA = TA.transitive_reduction()
  # to remove unnecessary edges added in last
  # two steps, if any.

  return TA

#Construct Generic Subtyping Product (GSP)
def GSP(subclassing, lngc, TAs, lbl_fn):
  # main step
  S=DiGraph.cartesian_product(subclassing,TAs)

  lngcc = map(lambda ngc: filter(lambda(c,_):
                c==ngc, S.vertices()), lngc)
  # lngcc is list of non-generic class clusters

  # merge the clusters
  map(lambda cc: S.merge_vertices(cc), lngcc)

  S = S.transitive_reduction()
  S.relabel(lambda (c,ta): c if (c in lngc)
                           else lbl_fn(c,ta))

  return S

def wty(c,wta):
  return c+WLP+wta+WRP

def WildcardsSubtyping(subclassing, lngc,
                       FN_Prfx, num_iter):
  #Definition of S0 (initial S)
  S0=subclassing.copy()
  S0.relabel(lambda c: c if (c in lngc) else
                       wty(c,W))

  S = S0
  lst = [S0]

  for i in [1..num_iter]:
    TA = TAs(S)

    # main step
    S = GSP(subclassing, lngc, TA, wty)

    lst.append(TA)
    lst.append(S)
 
  # Repeat as needed.
  # S = F(S) ... final (infinite) subtyping
  #  relation is soln of this eqn (a fixed
  #  point).

  return lst
\end{lstlisting}
(Note: The code presented in Appendix~A of~\citep{AbdelGawad2018c}
builds on and makes use of the SageMath code presented here.)
\end{document}